\documentclass[reprint,pdftex]{revtex4-1}
\usepackage{amsmath,amssymb,amsthm,upgreek}
\usepackage[dvips]{graphicx,color}
\usepackage{epstopdf}
\usepackage[pdftex,colorlinks]{hyperref}
\numberwithin{equation}{section}
\newcommand{\sevec}{\mathbf{Z}}
\begin{document}

\title{The strong-coupling master equation of quantum open systems}
\author{C.~H.~Fleming}
\affiliation{Joint Quantum Institute and Department of Physics, University of Maryland, College Park, Maryland 20742}
%\author{B.~L.~Hu}
%\affiliation{Joint Quantum Institute and Department of Physics, University of Maryland, College Park, Maryland 20742}
\date{\today}

\begin{abstract}
In this paper we demonstrate how to generate the strong-coupling master equations for open quantum systems of continuous variables.
These are the dissipative master equations of quantum Brownian particles for which the environmental noise is stronger than other system forces.
Our strong-coupling master equations are very different from other so-called ``strong-coupling'' master equations (e.g. the quantum Smoluchowski equation) which are perturbing off a limit in which the system energy is taken to be perturbative and thus the dynamics is principally Markovian.
Such approximations also require the system mass to be asymptotically large (even as compared to the ratio of noise and induced system frequencies) and thus they do not fully categorize the regime of what one might consider to be strong coupling.
Our master equations are highly non-Markovian and radically different for different system potentials, admitting no apparent generic form.
This result is quite exciting as it brings forth a new regime for theoretical exploration: the regime of strong noise and dissipation yet non-Markovian, such as strong coupling to a low-temperature environment with large 1/f fluctuations.
\end{abstract}

\maketitle

\tableofcontents

%%%%%%%%%
\section{Introduction}
An open system is obtained by coarse graining over the environmental degrees of freedom, 
and its dynamics is described by a master equation governing the reduced density matrix or marginal Wigner function.
The coarse-grained environment can act as a source of noise, dissipation, and decoherence;
thus its influence provides a microscopic model of dissipative quantum mechanics more general than Markovian (white-noise) models which can be constructed more phenomenologically.
Exact master equations for the stochastic dynamics of open quantum systems are, in general, beyond the reach of the simple theorist.
The canonical regime of interest for nonlinear quantum open systems has been one of weak coupling between the system and environment.
The resultant perturbative master equations for the dynamics of the reduced system are well known and can be derived in a variety of ways \cite{Kampen97,Breuer01,Strunz04}.
At zeroth-order the dynamics are that of the uncoupled system, and further orders consistently introduce noise and backaction from the environment.

At the other end of the spectrum, there is also great interest in the strong-damping regime.
Naive expectations of behavior in this regime appeal to the limit in which the system Hamiltonian is taken to be vanishing and the system is primarily driven by the system-environment interaction.
For a single (collective) coupling to the environment, or univariate noise, this produces a rapid decoherence in the basis of the system coupling operator.
Ergo, for a strong position coupling to the environment, it is thought that the system should decohere into a position distribution.
However, this limit has questionable applicability to small mechanical systems.
Large noise fluctuations can generate large kinetic energies, which are then incompatible with the limit of vanishing system energy.

In this work we derive from first principles the master equation of a quantum system primarily driven by noise and dissipation.
What we take to be perturbative is not the entire system Hamiltonian as with the ``strong-coupling'' \emph{quantum Smoluchowski equation}\cite{Ankerhold01,Coffey07} but only the system potential, or nonlinearities thereof.
The result is a class of master equations which are highly non-Markovian and seem to admit no generic form.
Instead we have a procedure for generating said master equations through an iterative application of noise averaging and integration.

%%%%%%%%%%
\section{Open system review}
\subsection{Classical-like Brownian trajectories}
Let us consider the canonical non-linear Brownian motion problem:
a Brownian particle (the system) with phase-space coordinates  ${x}$, ${p}$ and potential $V({x})$
bilinearly coupled to an environment consisting of many oscillators with coordinates ${x}_k$ and ${p}_k$.
Linear coupling to a bath of oscillators provides a general model of Gaussian noise~\cite{Feynman63}.
Although we take the dynamical equations to be of the classical variety, we will choose a quantum initial state for the environment.
The reason for this will become clear when we derive our perturbative formalism.

We can write the full system + environment Hamiltonian
\begin{align}
{H}_\mathrm{C} =&\, \frac{1}{2m}{p}^2 + V({x}) \label{eq:HC} \\
&+ \sum_k \left[ \frac{1}{2m_k}{p}_k^2 + \frac{1}{2}m_k\omega_k^2 \left({x}_k-\frac{c_k {x}}{m_k\omega_k^2}\right)^2 \right] \, , \nonumber
\end{align}
with $c_k$ the coupling strength and where the system-environment coupling has been inserted in the bath potential as a method of renormalization:
in this manner the phenomenological linear spring constant of the system is the same for any system-environment coupling $c_k$.
The system and environment are taken to be initially uncorrelated and uncoupled,
with the environment initially being in its free (quantum) thermal state.
The interaction is switched on instantaneously via $c_k$ at the initial time.
The equations of motion for the system are then
\begin{align}
\dot{{x}}(t) &= \frac{1}{m} {p}(t) \, , \\
\dot{{p}}(t) &= -V'[{x}(t)] -m \delta \omega_\mathrm{R}^2 {x}(t) + \sum_k c_k \, {x}_k(t) \, ,
\end{align}
and the equations of motion for the environment are
\begin{align}
\dot{{x}}_k(t) &= \frac{1}{m_k} {p}_k(t) \, , \\
\dot{{p}}_k(t) &= -m_k \omega_k^2 \, {x}_k(t) + c_k \, {x}(t) \, .
\end{align}
For each equation the momentum can be substituted by position to obtain
\begin{align}
m \,\ddot{{x}}(t) + V'[{x}(t)] + m \delta \omega_\mathrm{R}^2 \, {x}(t) &= \sum_k c_k \, {x}_k(t) \, , \\
m_k \, \ddot{{x}}_k(t) + m_k \omega_k^2 \, {x}_k(t) &= c_k \, {x}(t) \, .
\end{align}
Additionally, the environment positions can be solved in terms of the system position
\begin{align}
{x}_k(t) &= m_k \dot{f}_i(t) \, {x}_k(0) + f_k(t) \, {p}_k(0) + c_k f_k(t) * {x}(t) \, , \\
f_k(t) &\equiv  \frac{\sin(\omega_k t)}{m_k \omega_k} \, ,
\end{align}
where $*$ denotes the Laplace convolution $A(t)*B(t) = \int_0^t d\tau \, A(t\!-\!\tau) \, B(\tau)$ and $f_k(t)$ is the free Green's function of an environment oscillator.
This expression can be substituted back into the equation of motion for the system position,
providing the quantum Langevin equation
\begin{equation}
m \, \ddot{{x}}(t) + V'[{x}(t)] + 2 m \gamma(t) * {x}(t) + 2 m \gamma(t) \, {x}(0) = {\xi}(t) \, , 
\end{equation}
where the positive-definite, stationary damping kernel $\gamma(t,\tau) = \gamma(t\!-\!\tau)$ and Gaussian noise process ${\xi}(t)$ are respectively given by
\begin{align}
\gamma(t,\tau) & \equiv  \frac{1}{2m} \sum_k \frac{c_k^2}{m_k \omega_k^2} \cos[\omega_k(t\!-\!\tau)] \, ,  \\
{\xi}(t) &\equiv \sum_k c_k \left[ m_k \dot{f}_i(t) \, {x}_k(0) + f_k(t) \, {p}_k(0) \right] \, .
\end{align}
The stationary noise kernel is defined
\begin{equation}
\nu(t,\tau) \,=\, \left\langle {\xi}(t) \, {\xi}(\tau) \right\rangle_{{\xi}} \,=\, \nu(\!t-\!\tau) \,
\end{equation}
and is related to damping by the (quantum) fluctuation-dissipation relation
\begin{eqnarray}
\tilde{\nu}(\varepsilon) &=& m \, \tilde{\gamma}(\varepsilon) \, \varepsilon \coth\!\left( \frac{\varepsilon}{2T} \right) \, , \label{eq:nu}
\end{eqnarray}
where $\tilde{\gamma}(\varepsilon) = \int_{-\infty}^{+\infty} \! dt \, e^{-\imath \varepsilon t} \, \gamma(t)$.
Essentially, the damping kernel (equivalently the spectral-density function) and temperature completely characterize Gaussian, thermal noise.

%%%%%%%%%
\subsubsection{Linear systems} \label{sec:Phi}
For a linear system the Langevin equation takes the classical form
\begin{equation}
m \, \ddot{{x}}(t) + m \omega^2 \, {x}(t) + 2 m \gamma(t) * {x}(t) + 2 m \gamma(t) \, {x}(0) = {\xi}(t) \, ,
\end{equation}
and the solutions are given by
\begin{equation}
{x}(t) = m \dot{g}(t) \, {x}(0) + g(t) \, {p}(0) + g(t) * {\xi}(t) \, ,
\end{equation}
where the Green's function $g(t)$ is most easily determined in the Laplace domain as
\begin{eqnarray}
\hat{g}(s) &=& \frac{\frac{1}{m}}{s^2 + \omega^2 + 2 s \hat{\gamma}(s)} \, ,
\end{eqnarray}
where $\hat{g}(s) = \int_0^\infty dt \, e^{-st} \, g(t)$.
Defining the phase-space coordinates $\mathbf{z}$ and homogeneous system propagator $\boldsymbol{\Phi}(t)$ as
\begin{eqnarray}
\mathbf{z}(t) &\equiv& [ {x}(t) , {p}(t) ]^\mathrm{T} \, , \\
\boldsymbol{\Phi}(t) &\equiv& \left[ \begin{array}{cc} m\dot{g}(t) & g(t) \\ m^2 \ddot{g}(t) & m \dot{g}(t) \end{array} \right] \, , \label{eq:Phi}
\end{eqnarray}
with free environment propagator
\begin{eqnarray}
\boldsymbol{\upphi}_{\!k}(t) &\equiv& \left[ \begin{array}{cc} m_k\dot{f}_k(t) & f_k(t) \\ m_k^2 \ddot{f}_k(t) & m \dot{f}_k(t) \end{array} \right] \, , \label{eq:phi}
\end{eqnarray}
we can express the system phase-space solutions
\begin{align}
\mathbf{z}(t) &= \boldsymbol{\Phi}(t) \, \mathbf{z}(0) + \boldsymbol{\Phi}(t) * \sum_k \mathbf{c}_k \, \boldsymbol{\upphi}_{\!k}(t) \, \mathbf{z}_k(0) , \label{eq:q(t)normal} \\
\mathbf{c}_k &\equiv \left[ \begin{array}{cc} 0 & 0 \\ 0 & c_k \end{array} \right] \, ,
\end{align}
and the homogeneous system propagator $\boldsymbol{\Phi}(t)$ can be identified as the system-system portion of the full system-environment propagator.
Similarly the environment solutions can be expressed
\begin{align}
\mathbf{z}_k(t) =&\, \boldsymbol{\upphi}_{\!k}(t) * \mathbf{c}_k \, \boldsymbol{\Phi}(t) \, \mathbf{z}(0) \\
&+ \sum_j \left[ \delta_{kj} \delta(t) + \boldsymbol{\upphi}_{\!k}(t) * \mathbf{c}_k \, \boldsymbol{\Phi}(t) \, \mathbf{c}_j \right] *  \boldsymbol{\phi}_j(t) \, \mathbf{z}_j(0) \, . \nonumber
\end{align}

The stochastic description extends to our phase-space representation with stochastic process $\boldsymbol{\Xi}(t) = [0,{\xi}(t)]^\mathrm{T}$ in phase-space coordinates, such that
\begin{eqnarray}
\mathbf{z}(t) &=& \boldsymbol{\Phi}(t) \, \mathbf{z}(0) + \boldsymbol{\Phi}(t) * \boldsymbol{\Xi}(t) \, , \label{eq:q(t)}
\end{eqnarray}
where the Gaussian noise correlation is now given by
\begin{equation}
\boldsymbol{\nu}(t,\tau) = \left\langle \boldsymbol{\Xi}(t) \, \boldsymbol{\Xi}^\mathrm{T}(\tau) \right\rangle_{\!\!\boldsymbol{\Xi}} = \left[ \begin{array}{cc} 0 & 0 \\ 0 & \nu(t,\tau) \end{array} \right] \, , \label{eq:nuq}
\end{equation}
where $\nu(t,\tau)$ is the noise correlation of ${\xi}(t)$ given by Eq.~\eqref{eq:nu}.

%%%%%%%%%%%
\subsection{Linear quantum Brownian motion} \label{sec:QBML}
For a linear system, the master equation \cite{HPZ92} of the reduced density matrix $\boldsymbol{\rho}$
can also be represented in the phase-space representation \cite{QBM} as a Fokker-Plank equation of the reduced Wigner function $W\!(\mathbf{z},t)$,
given by
\begin{align}
\frac{d}{dt} W\!(\mathbf{z},t) &= \boldsymbol{\mathcal{L}}(t) \,  W\!(\mathbf{z},t) \, , \label{eq:LQBM} \\
\boldsymbol{\mathcal{L}}(t) &\equiv \boldsymbol{\nabla}_{\!\!\mathbf{z}}^\mathrm{T} \, \boldsymbol{\mathcal{H}}(t) \, \mathbf{z} + \boldsymbol{\nabla}_{\!\!\mathbf{z}}^\mathrm{T} \, \mathbf{D}(t) \, \boldsymbol{\nabla}_{\!\!\mathbf{z}} \, . \label{eq:QBML}
\end{align}
The Fokker-Planck equation takes a classical form except that the diffusion coefficients describe quantum fluctuations.
The time-local homogeneous and diffusion coefficients are given by
\begin{align}
\boldsymbol{\mathcal{H}}(t) &= -\dot{\boldsymbol{\Phi}}(t) \, \boldsymbol{\Phi}^{-1}(t) \, , \\
\mathbf{D}(t) &= \frac{1}{2} \left\{ \boldsymbol{\mathcal{H}}(t) \, \boldsymbol{\sigma}_{\!T}(t) + \boldsymbol{\sigma}_{\!T}(t) \, \boldsymbol{\mathcal{H}}^\mathrm{T}(t) + \dot{\boldsymbol{\sigma}}_T(t) \right\} \, ,
\end{align}
where the homogeneous propagator $\boldsymbol{\Phi}(t)$ is given by Eq.~\eqref{eq:Phi}
and the thermal covariance is
\begin{equation}
\boldsymbol{\sigma}_{\!T}(t) \equiv \int_0^t \!\! d\tau_1 \! \int_0^t \!\! d\tau_2 \, \boldsymbol{\Phi}(t-\tau_1) \, \boldsymbol{\nu}(\tau_1 , \tau_2) \, \boldsymbol{\Phi}^\mathrm{T}(t-\tau_2) \, .
\end{equation}
This is a special case of the two-time thermal covariance
\begin{equation}
\boldsymbol{\sigma}_{\!T}(t_1,t_2) \equiv \int_0^{t_1} \!\!\! d\tau_1 \! \int_0^{t_2} \!\!\! d\tau_2 \, \boldsymbol{\Phi}(t_1-\tau_1) \, \boldsymbol{\nu}(\tau_1 , \tau_2) \, \boldsymbol{\Phi}^\mathrm{T}(t_2-\tau_2) \, , \label{eq:sigmaT}
\end{equation}
which is the thermal contribution to the two-time correlation function
\begin{equation}
\left\langle \mathbf{z}(t_1) \, \mathbf{z}^\mathrm{T}(t_2) \right\rangle_{\boldsymbol{\Xi}} = \boldsymbol{\Phi}(t_1) \, \boldsymbol{\sigma}_0 \, \boldsymbol{\Phi}^\mathrm{T}(t_2) + \boldsymbol{\sigma}_{\!T}(t_1,t_2) \, ,
\end{equation}
given the initial correlation $\boldsymbol{\sigma}_0 \equiv \left\langle \mathbf{z}(0) \, \mathbf{z}^\mathrm{T}(0) \right\rangle$.
The first contribution is thus homogeneous and dissipative.

%%%%%%%%%
\section{A new perturbative formalism}
\subsection{Classical Dynamics}
Let $\mathbf{z} = (x,p)$ and $\mathbf{z}_k = (x_k,p_k)$ denote the individual system and environment phase-space coordinates respectively.
Let $\sevec = (\mathbf{z},\mathbf{z}_1,\mathbf{z}_2,\cdots)$ denote the collective system + environment phase-space coordinates.
The classical dynamics of phase-space distributions is such that
\begin{eqnarray}
W\![\sevec(t),t] &=& W\![\sevec(0),0] \, ,
\end{eqnarray}
where $\sevec(t)$ is a classical trajectory with initial conditions $\sevec(0)$.
Therefore the classical propagator $\mathbf{G}_\mathrm{C}(t_1,t_2) : W\!(\sevec,t_2) \to W\!(\sevec,t_1)$ must act upon arbitrary distributions $d$ such that
\begin{eqnarray}
\mathbf{G}_\mathrm{C}(t_1,t_2) \, d[\sevec(t_1)] &=&  d[\sevec(t_2)] \, . \label{eq:Gclassical}
\end{eqnarray}
The classical trajectories are determined by classical equations of motion of the form
\begin{eqnarray}
\frac{d}{dt} \sevec(t) &=& \mathbf{F}[\sevec(t),t] \, ,
\end{eqnarray}
and correspondingly, the classical Fokker-Plank equation is given by
\begin{eqnarray}
\frac{\partial}{\partial t} W\!(\sevec,t) &=& - \mathbf{F}(\sevec,t)^\mathrm{T} \, \boldsymbol{\nabla}_{\!\!\sevec} \, W\!(\sevec,t) \, .
\end{eqnarray}
In contrast, the quantum master equation is not generally linear in derivatives, thus not admitting trajectories or ``characteristics''.
For polynomial potentials beyond the quadratic order, the quantum equation contains higher-order derivatives accompanied by corresponding factors of $\hbar$.
In general the time-translation generator is nonlocal in the phase-space coordinates coordinates \cite{Hillery84}.

%%%%%%%%%%
\subsection{Quantum dynamics in the characteristics picture}
We now examine the quantum dynamics of the combined system + environment generated by a Hamiltonian, Eq.~\eqref{eq:HC}, which yields unitary evolution.
In the quantum phase-space representation, the Wigner function obeys 
\begin{eqnarray}
\frac{\partial}{\partial t} W\![\sevec,t] &=& \mathcal{L}_{\sevec}[\boldsymbol{\nabla}_{\!\!\sevec},\sevec,t] \, W\![\sevec,t] \, , 
\end{eqnarray}
where $\mathcal{L}_{\sevec}$ denotes the unitary generator of the system + environment (compare with Eq.~\eqref{eq:LQBM} of the open system).
Let us inspect the dynamics of quantum states along some curves $\sevec(t)$ which are not assumed to be proper characteristics.
\begin{align}
\frac{d}{dt} W\![\sevec(t),t] =&\, \left[\frac{d}{dt} \sevec(t)\right]^\mathrm{T} \!\! \boldsymbol{\nabla}_{\!\!\sevec(t)} W\![\sevec(t),t] \\
& + \mathcal{L}_{\sevec}[\boldsymbol{\nabla}_{\!\!\sevec(t)},\sevec(t),t] \, W\![\sevec(t),t] \, . \nonumber
\end{align}
If these curves have \emph{some} first-order equations of motion governed by the ``forces'' $\mathbf{F}(\sevec,t)$
then we can express the dynamics along these curves as
\begin{align}
\frac{d}{dt} W\![\sevec(t),t] &= \delta\mathcal{L}[\boldsymbol{\nabla}_{\!\!\sevec(t)},\sevec(t),t] \, W\![\sevec(t),t] \, , \label{eq:WCL} \\
\boldsymbol{\delta\!\mathcal{L}}(t) & \equiv  \boldsymbol{\mathcal{L}}_{\sevec}(t) + \mathbf{F}(\sevec,t)^\mathrm{T} \, \boldsymbol{\nabla}_{\!\!\sevec} \, . \label{eq:dL}
\end{align}
The key point is that we will utilize curves $\sevec(t)$ such that $\boldsymbol{\delta\!\mathcal{L}}(t)$ is ordinarily a \emph{system} operator
and then calculate the corresponding perturbative open-system master equation for the marginal distribution $W\!(\mathbf{z},t)$,
which is equivalent to the reduced density matrix.

Any Gaussian influence acting on the system can be modeled via linear coupling to a linear environment,
and the dynamical contributions of all such terms are first-order (dynamically classical) in the phase-space representation.
Therefore we can always transform to a classically-evolving coordinate system, along which the dynamics of the environment are effectively integrated out.
Moreover, the system Hamiltonian used for the classical curves is irrelevant for the specific purpose of integrating out the environment dynamics.
We do not have to use the true system Hamiltonian for our characteristics,
but for the purposes of perturbation theory we do want $\boldsymbol{\delta\!\mathcal{L}}(t)$ to be small.
Therefore we generally want the classical dynamics to reproduce as much of the quantum dynamics as possible, allowing for much cancellation in Eq.~\eqref{eq:dL}.
Essentially, the noise average of Eq.~\eqref{eq:WCL} is the nonlinear generalization of the stochastic description of QBM found in Ref.~\cite{CRV01}.
Whereas for linear systems the stochastic trajectories $\mathbf{z}(t)$ immediately provide the solution,
here they only serve as an evolving coordinate system which encapsulates all effects of the environment.

%%%%%%%%%%
\subsubsection{Perturbation along characteristics}
The integral equation of motion corresponding to Eq.~\eqref{eq:WCL} is
\begin{align}
W\![\sevec(t),t] =&\, W\![\sevec(0),0] \\
& + \int_0^t \!\! d\tau \, \delta\mathcal{L}[\boldsymbol{\nabla}_{\!\!\sevec(\tau)},\sevec(\tau),\tau] \, W\![\sevec(\tau),\tau] \, , \nonumber
\end{align}
which is amenable to perturbation via a Neumann series yielding the perturbative solutions
\begin{align}
W_{\!0}[\sevec(t),t] &= W\![\sevec(0),0] \, , \\
W_{\!1}[\sevec(t),t] &=  \int_0^t \!\! d\tau \, \delta\mathcal{L}[\boldsymbol{\nabla}_{\!\!\sevec(\tau)},\sevec(\tau),\tau] \, W\![\sevec(0),0] \, .
\end{align}
and so on, where $W\![\sevec,t] = \sum_{n=0}^\infty W_n[\sevec,t]$
and $W_n[\sevec,t] = \mathcal{O}(\delta\mathcal{L}^n)$.
Then we can apply the classical propagator of Eq.~\eqref{eq:Gclassical} to transform back to the (initial) domain coordinates.
\begin{align}
W_{\!n}[\sevec(0),t] &= \mathbf{G}_\mathrm{C}(t,0) \, W_{\!n}\![\sevec(t),t] \, .
%W_0[\sevec(0),t] &= \mathbf{G}_\mathrm{C}(t,0) \, W\![\sevec(0),0] \, , \\
%W_1[\sevec(0),t] &=  \int_0^t \!\! d\tau \, \mathbf{G}_\mathrm{C}(t,\tau) \, \delta\mathcal{L}[\boldsymbol{\nabla}_{\!\!\sevec(0)},\sevec(0),\tau] \, \mathbf{G}_\mathrm{C}(\tau,0) \, W\![\sevec(0),0] \, . 
\end{align}
Next we assume the initial state of the system + environment to be an uncorrelated product of marginal distributions,
$W\!(\sevec,0) = W\!(\mathbf{z},0) \prod_k W\!(\mathbf{z}_k,0)$.
Upon tracing over the environment we obtain the perturbative open-system propagator
\begin{align}
\mathbf{G}_0(t) &= \left\langle \mathbf{G}_\mathrm{C}(t,0) \right\rangle_\mathrm{E} \, , \label{eq:G0} \\
\mathbf{G}_1(t) &= \int_0^t \!\! d\tau \left\langle \mathbf{G}_\mathrm{C}(t,\tau) \, \boldsymbol{\delta\!\mathcal{L}}(t) \, \mathbf{G}_\mathrm{C}(\tau,0) \right\rangle_\mathrm{E} \, .
\end{align}
and so on for $\mathbf{G}_n(t)$ such that 
\begin{equation}
\mathbf{G}(t) = \sum_{n=0}^\infty \mathbf{G}_n(t) : W\!(\mathbf{z},0) \to W\!(\mathbf{z},t) \, ,
\end{equation}
and where the operator
\begin{equation}
\boldsymbol{\delta\!\mathcal{L}}(t) \,=\, \delta\mathcal{L}[\boldsymbol{\nabla}_{\!\!\sevec},\sevec,t] \,=\, \delta\mathcal{L}[\boldsymbol{\nabla}_{\!\!\mathbf{z}},\mathbf{z},t]\, ,
\end{equation}
is always taken to be exclusively a system operation.
Perturbative solutions of this form are inherently secular in time and do not respect Lie group symmetries such as unitary or completely-positive semi-group evolution.
Instead we will only use this expansion to obtain the perturbative generators.
Using ordinary perturbation theory, our open-system master equation is then
\begin{align}
\boldsymbol{\mathcal{L}}_0(t) &= \left[ \frac{d}{dt} \left\langle \mathbf{G}_\mathrm{C}(t,0) \right\rangle_\mathrm{E} \right] \left\langle \mathbf{G}_\mathrm{C}(t,0) \right\rangle_\mathrm{E}^{-1} \, , \label{eq:L0} \\
\boldsymbol{\mathcal{L}}_1(t) &= \boldsymbol{\delta\!\mathcal{L}}(t) + \int_0^t \!\! d\tau \left\{ \frac{d}{dt} -\mathrm{Ad}[\boldsymbol{\mathcal{L}}_0(t)] \right\} \underline{\boldsymbol{\delta\!\mathcal{L}}}(\tau,t) \, , \label{eq:L1} \\
\underline{\boldsymbol{\delta\!\mathcal{L}}}(\tau,t) &\equiv \left\langle \mathbf{G}_\mathrm{C}(t,\tau) \, \boldsymbol{\delta\!\mathcal{L}}(\tau) \, \mathbf{G}_\mathrm{C}(\tau,0) \right\rangle_{\mathrm{E}} \left\langle \mathbf{G}_\mathrm{C}(t,0) \right\rangle_\mathrm{E}^{-1} \, ,
\end{align}
and so on, where $\boldsymbol{\mathcal{L}}(t) = \sum_{n=0}^\infty \boldsymbol{\mathcal{L}}_n(t)$ is the open-system Liouvillian or time-evolution generator,
such that
\begin{eqnarray}
\frac{\partial}{\partial t} W\![\mathbf{z},t] &=& \boldsymbol{\mathcal{L}}(t) \, W\![\mathbf{z},t] \, .
\end{eqnarray}

The lowest-order propagator $\mathbf{G}_0(t)$, given by Eq.~\eqref{eq:G0}, is the evolution operator for the classical Brownian motion problem, but with quantum noise, making it potentially exact for linear systems.
$\boldsymbol{\mathcal{L}}_0(t)$, given by Eq.~\eqref{eq:L0}, is the corresponding Fokker-Plank equation but with quantum noise, e.g. the HPZ equation \cite{HPZ92}.
Essentially, if one can solve the classical Brownian motion problem (with quantum fluctuations)
then one can determine the quantum corrections due to the combination of nonlinearities and noise.
If one is in the classical, linear, Markovian or noiseless regime, then all perturbative corrections can vanish in this formalism given the appropriate characteristics to perturb from.

Our hierarchy of approximation schemes is therefore not only determined by the order to which we calculate the master equation, the $n$ of $\boldsymbol{\mathcal{L}}_n$,
but also the amount of detail about our system captured in the characteristics we use.
%, as described by Table.~\ref{tab:char}.
%\begin{table}[h!]
%\begin{tabular}{|c|c|c|}
%\hline
%\textbf{Characteristics} & \textbf{System Potential} & \textbf{Regime} \\
%\hline
%Damped Free Particle		& $0$						& Strong Damping \\
%Damped Oscillator			& $\frac{1}{2}m \omega^2 x^2$	& Quasi-Linear Feedback Dynamics \\
%Damped Classical Particle	& $V(x)$					& Semi-Classical Feedback Dynamics \\
%\hline
%\end{tabular}
%\caption{Perturbative schemes organized by their characteristics. \label{tab:char}}
%\end{table}
Perturbing off the damped (but otherwise) free characteristics is strong-damping perturbation as $\boldsymbol{\mathcal{L}}_n$ is $\mathcal{O}(V^n)$.
Perturbing off of the damped oscillator characteristics is a kind of quasi-linear feedback approximation.
And perturbing off of the full classical characteristics is a kind of semi-classical feedback approximation.
At least formally, one might also imagine an extension beyond the classical characteristics into the quantum characteristics \cite{Krivoruchenko07},
possibly with an additional (partial) semi-classical expansion therein.
Each class of characteristics encompasses the previous.

Note that the system operation $\boldsymbol{\delta\!\mathcal{L}}(t)$ is fully retained in the first-order correction $\boldsymbol{\mathcal{L}}_1(t)$;
only its effect through the environment (back-action or feedback) is approximate.
To zeroth order, the environment perceives the system evolution as being linear or classical and its feedback upon the environment is therefore approximate in that manner.
At first order we pick up the full nonlinearity of the system in $\boldsymbol{\delta\!\mathcal{L}}(t)$ along with feedback corrections from the time integral in Eq.~\eqref{eq:L1}.
Essentially $\underline{\boldsymbol{\delta\!\mathcal{L}}}(\tau,t)$ is a two-time and open-system generalization of $\boldsymbol{\delta\!\mathcal{L}}(t)$,
such that $\underline{\boldsymbol{\delta\!\mathcal{L}}}(t,t) = \boldsymbol{\delta\!\mathcal{L}}(t)$ and $\underline{\boldsymbol{\delta\!\mathcal{L}}}(0,t) = \mathbf{G}_0(t) \, \boldsymbol{\delta\!\mathcal{L}}(t) \,  \mathbf{G}_0^{-1}(t)$,
which is integrated over the past in order to compensate for what is missed in the lower-order approximation.

This is an analogous formalism to the more common weak-damping perturbation wherein the 
system and environment are uncoupled at zeroth order.
In weak-damping perturbation the higher-order generators involve integrals over the past which contain the free propagators acting upon the system-environment interaction.
Here in strong-damping perturbation the higher-order generators involve integrals over the past which contain the approximate (yet interacting) propagators acting upon the system nonlinearities.

%%%%%%%%%%%%
\section{Linear back-action} \label{sec:LinearBack}
Perturbing off a linear system + environment is the most straightforward application of this formalism.
This includes both the quasi-linear feedback and strong-coupling regimes.
In evaluating the first-order Liouville operator in Eq.~\eqref{eq:L1}, we aim to calculate the two-time and open-system operator
\begin{equation}
\underline{\boldsymbol{\delta\!\mathcal{L}}}(\tau,t) = \left\langle \mathbf{G}_\mathrm{C}(t,\tau) \, \boldsymbol{\delta\!\mathcal{L}}(\tau) \, \mathbf{G}_\mathrm{C}(\tau,0) \right\rangle_{\mathrm{E}} \left\langle \mathbf{G}_\mathrm{C}(t,0) \right\rangle_\mathrm{E}^{-1} \, ,
\end{equation}
for all ordinary system operators $\boldsymbol{\delta\!\mathcal{L}}(t)$.
This turns out to be a complicated procedure and therefore there will be no simple strong-coupling master equation which we can write down for an arbitrary system potential.
Instead one must apply the following procedures iteratively and work out the master equation for specific system potentials.

%%%%%%%%%%%%%
\subsection{Evaluation of two-time open-system operators} \label{eq:LinearRules}
\subsubsection{Transformation of Derivatives}
As a simple example, let us consider the reduced (linear) derivative operator
\begin{align}
\boldsymbol{\delta\!\mathcal{L}} &= \boldsymbol{\nabla}_{\!\!\mathbf{z}}^\mathrm{T} , \\
\underline{\boldsymbol{\delta\!\mathcal{L}}}(\tau,t) &= \left\langle \mathbf{G}_\mathrm{C}(t,\tau) \, \boldsymbol{\nabla}_{\!\!\mathbf{z}}^\mathrm{T} \, \mathbf{G}_\mathrm{C}(\tau,0) \right\rangle_{\!\!\mathrm{E}} \left\langle \mathbf{G}_\mathrm{C}(t,0) \right\rangle_\mathrm{E}^{-1} \, .
\end{align}
First we move all system derivatives to the left by performing the transformation
\begin{align}
& \underline{\boldsymbol{\delta\!\mathcal{L}}}(\tau,t) = \\
& \left\langle \left[ \mathbf{G}_\mathrm{C}(t,\tau) \, \boldsymbol{\nabla}_{\!\!\mathbf{z}}^\mathrm{T} \,  \mathbf{G}_\mathrm{C}(\tau,t) \right] \mathbf{G}_\mathrm{C}(t,0) \right\rangle_{\!\!\mathrm{E}} \left\langle \mathbf{G}_\mathrm{C}(t,0) \right\rangle_\mathrm{E}^{-1} \, . \nonumber
\end{align}
System derivatives transform with the transpose of the system + environment propagator
\begin{align}
\boldsymbol{\nabla}_{\!\!\mathbf{z}}^\mathrm{T} \to \underbrace{ \boldsymbol{\nabla}_{\!\!\mathbf{z}}^\mathrm{T} \, \boldsymbol{\Phi}(t-\tau) }_{\mathrm{system}} + \sum_k  \underbrace{ \boldsymbol{\nabla}_{\!\!\mathbf{z}_k}^\mathrm{T} \, [ \boldsymbol{\Phi} * \mathbf{c}_k \, \boldsymbol{\upphi}_{\!k}](t-\tau) }_{\mathrm{environment}} \, , 
\end{align}
where we have used the system portion of the full system + environment propagation detailed in Sec.~\ref{sec:Phi}.
%The system-system propagator $\boldsymbol{\Phi}(t-\tau)$, which is also the homogeneous system propagator, is not itself time-translation invariant.
Homogeneous system derivatives pass through the noise average,
whereas noise derivatives will be shown to trivially vanish in the noise average.
Integrating over environment derivatives results in only boundary terms which should limit to zero.
E.g.
\begin{equation}
\int_{-\infty}^{+\infty} \!\!\! dx_k \frac{\partial}{\partial x_k} W\!(x_k) \,=\, \left. W\!(x_k) \right|_{-\infty}^{+\infty} \,=\, 0 \, .
\end{equation}
Therefore our reduced derivative is simply
\begin{eqnarray}
\underline{\boldsymbol{\delta\!\mathcal{L}}}(\tau,t) &=& \boldsymbol{\nabla}_{\!\!\mathbf{z}}^\mathrm{T} \, \boldsymbol{\Phi}(t-\tau) \, , \label{eq:D(tau,t)}
\end{eqnarray}
when left of all other operations.

%%%%%%%%%
\subsubsection{Transformation of coordinates}
As a simple example, let us consider the reduced (linear) coordinate operator
\begin{align}
\boldsymbol{\delta\!\mathcal{L}} &= \mathbf{z} \, , \\
\underline{\boldsymbol{\delta\!\mathcal{L}}}(\tau,t) &= \left\langle \mathbf{G}_\mathrm{C}(t,\tau) \, \mathbf{z} \, \mathbf{G}_\mathrm{C}(\tau,0) \right\rangle_{\mathrm{E}} \left\langle \mathbf{G}_\mathrm{C}(t,0) \right\rangle_\mathrm{E}^{-1} \, .
\end{align}
First we move our system coordinates to the right by considering the transformation
\begin{align}
& \underline{\boldsymbol{\delta\!\mathcal{L}}}(\tau,t) = \\
& \left\langle \mathbf{G}_\mathrm{C}(t,0) \left[ \mathbf{G}_\mathrm{C}(0,\tau) \, \mathbf{z} \, \mathbf{G}_\mathrm{C}(\tau,0) \right] \right\rangle_{\mathrm{E}} \left\langle \mathbf{G}_\mathrm{C}(t,0) \right\rangle_\mathrm{E}^{-1} \, . \nonumber
\end{align}
System coordinates transform with the inverse of the propagator \eqref{eq:q(t)} and so we have
\begin{align}
& \underline{\boldsymbol{\delta\!\mathcal{L}}}(\tau,t) = \label{eq:Qrot} \\
& \left\langle \mathbf{G}_\mathrm{C}(t,0) \left[ \boldsymbol{\Phi}(\tau) \, \mathbf{z} + \boldsymbol{\Phi}(\tau) * \boldsymbol{\Xi}(\tau) \right] \right\rangle_{\boldsymbol{\Xi}} \left\langle \mathbf{G}_\mathrm{C}(t,0) \right\rangle_{\boldsymbol{\Xi}}^{-1} \, . \nonumber
\end{align}
in the stochastic representation.
Homogeneous system coordinates can pass through the noise average,
but then exist sandwiched between open-system propagators.
%\begin{eqnarray}
%\underline{\boldsymbol{\delta\!\mathcal{L}}}(\tau,t) &=&  \left\langle \mathbf{G}_\mathrm{C}(t,0) \right\rangle_{\boldsymbol{\Xi}} \boldsymbol{\Phi}(\tau) \, \mathbf{z} \left\langle \mathbf{G}_\mathrm{C}(t,0) \right\rangle_{\boldsymbol{\Xi}}^{-1} \\
%&& +\left\langle \mathbf{G}_\mathrm{C}(t,0) \, \boldsymbol{\Phi}(\tau) * \boldsymbol{\Xi}(\tau) \, \right\rangle_{\boldsymbol{\Xi}} \left\langle \mathbf{G}_\mathrm{C}(t,0) \right\rangle_{\boldsymbol{\Xi}}^{-1} \nonumber \, .
%\end{eqnarray}
To simplify the homogeneous part, we construct the generic identity
\begin{align}
\mathbf{A} \, \mathbf{z} &= \mathbf{A} \, \mathbf{z} \left\langle \mathbf{G}_\mathrm{C}(t,0) \right\rangle_{\boldsymbol{\Xi}} \left\langle \mathbf{G}_\mathrm{C}(t,0) \right\rangle_{\boldsymbol{\Xi}}^{-1} \, ,
\end{align}
which evaluates to
\begin{align}
\left\langle \mathbf{G}_\mathrm{C}(t,0) \left[  \mathbf{A} \, \boldsymbol{\Phi}(t) \, \mathbf{z} + \mathbf{A} \, \boldsymbol{\Phi}(t) * \boldsymbol{\Xi}(t) \right\rangle_{\boldsymbol{\Xi}} \right] \left\langle \mathbf{G}_\mathrm{C}(t,0) \right\rangle_{\boldsymbol{\Xi}}^{-1} \, , \label{eq:GenId}
\end{align}
and then we choose $\mathbf{A} = \boldsymbol{\Phi}(\tau,t)$ so that the homogeneous part of Eq.~\eqref{eq:GenId} matches with the homogeneous part of Eq.~\eqref{eq:Qrot}.
Therefore we now have
\begin{align}
\underline{\boldsymbol{\delta\!\mathcal{L}}}(\tau,t) &=  \boldsymbol{\Phi}(\tau,t) \, \mathbf{z} \\
& - \left\langle \mathbf{G}_\mathrm{C}(t,0) \, \boldsymbol{\Phi}(\tau,t) \, \boldsymbol{\Phi}(t) * \boldsymbol{\Xi}(t) \right\rangle_{\boldsymbol{\Xi}} \left\langle \mathbf{G}_\mathrm{C}(t,0) \right\rangle_{\boldsymbol{\Xi}}^{-1} \nonumber \\
& +\left\langle \mathbf{G}_\mathrm{C}(t,0) \, \boldsymbol{\Phi}(\tau) * \boldsymbol{\Xi}(\tau) \, \right\rangle_{\boldsymbol{\Xi}} \left\langle \mathbf{G}_\mathrm{C}(t,0) \right\rangle_{\boldsymbol{\Xi}}^{-1} \nonumber \, .
\end{align}
This expression can also be written
\begin{align}
& \underline{\boldsymbol{\delta\!\mathcal{L}}}(\tau,t) = \boldsymbol{\Phi}(\tau,t) \, \mathbf{z} \\
&+ \left\langle \mathbf{G}_\mathrm{C}(t,0) \int_0^t \!\! d\tau' \, \boldsymbol{\Phi}_\mathrm{f}(\tau,\tau') \, \boldsymbol{\Xi}(\tau') \right\rangle_{\!\!\boldsymbol{\Xi}}  \left\langle \mathbf{G}_\mathrm{C}(t,0) \right\rangle_{\boldsymbol{\Xi}}^{-1} \, , \nonumber
\end{align}
where the $\boldsymbol{\Phi}_\mathrm{f}(\tau,\tau')$ is  the final-value propagator \cite{QBM},
and is given by 
\begin{align}
\boldsymbol{\Phi}_\mathrm{f}(\tau,\tau') & \equiv  \theta(\tau\!-\!\tau') \, \boldsymbol{\Phi}(\tau\!-\!\tau') - \boldsymbol{\Phi}(\tau,t) \, \boldsymbol{\Phi}(t\!-\!\tau') \, .
\end{align}
This would be the \emph{advanced propagator} for local dissipation
\begin{eqnarray}
\boldsymbol{\Phi}_\mathrm{adv}(\tau,\tau') & \equiv & - \theta(\tau'\!-\!\tau) \, \boldsymbol{\Phi}(\tau\!-\!\tau') \, ,
\end{eqnarray}
but for nonlocal dissipation no such propagator exists.

%%%%%%%%%%%%
\subsubsection{Noise averages}
Tracing over environment coordinates will yield moments of the noise.
Most easily, the environment coordinates can be identified with the stochastic process $\boldsymbol{\Xi}(t)$, a Gaussian noise process with noise correlation $\langle \boldsymbol{\Xi}(t) \, \boldsymbol{\Xi}^\mathrm{T}(\tau) \rangle_{\boldsymbol{\Xi}} = \boldsymbol{\nu}(t,\tau)$ given by Eq.~\eqref{eq:nuq}.
Therefore we can evaluate our noise averages with the help of Novikov's formula (functional integration by parts), e.g.
\begin{align}
& \left\langle \mathbf{G}_\mathrm{C}(t,0) \, \boldsymbol{\Xi}(\tau') \right\rangle_{\boldsymbol{\Xi}} = \label{eq:ex1} \\
& -\int_0^t \!\! d\tau'' \left\langle \mathbf{G}_\mathrm{C}(t,0) \, \boldsymbol{\nu}(\tau',\tau'') \left[ \frac{\delta \mathbf{z}(t)}{\delta \boldsymbol{\Xi}(\tau'')} \right]^\mathrm{\!T} \!\!  \boldsymbol{\nabla}_{\!\!\mathbf{z}(t)} \right\rangle_{\!\!\boldsymbol{\Xi}} \, , \nonumber
\end{align}
where from Eq.~\eqref{eq:q(t)}, the functional derivative must be
\begin{eqnarray}
\left[ \frac{\delta \mathbf{z}(t)}{\delta \boldsymbol{\Xi}(\tau'')} \right] &=& \boldsymbol{\Phi}(t\!-\!\tau'') \, ,
\end{eqnarray}
and so for calculating \eqref{eq:ex1} we have
\begin{align}
& \left\langle \mathbf{G}_\mathrm{C}(t,0) \, \boldsymbol{\Xi}(\tau') \right\rangle_{\boldsymbol{\Xi}} = \\
& - \int_0^t \!\! d\tau'' \boldsymbol{\nu}(\tau',\tau'') \, \boldsymbol{\Phi}^\mathrm{T}(t\!-\!\tau'') \boldsymbol{\nabla}_{\!\!\mathbf{z}} \left\langle \mathbf{G}_\mathrm{C}(t,0) \right\rangle_{\boldsymbol{\Xi}} \, , \nonumber
\end{align}
upon substitution and transforming the derivative through the propagator.
All together the co-rotating coordinate becomes
\begin{align}
\underline{\boldsymbol{\delta\!\mathcal{L}}}(\tau,t) &= \boldsymbol{\Phi}(\tau,t) \, \mathbf{z} \\
&- \int_0^t \!\! d\tau' \!\! \int_0^t \!\! d\tau'' \, \boldsymbol{\Phi}_\mathrm{f}(\tau,\tau') \, \boldsymbol{\nu}(\tau',\tau'') \, \boldsymbol{\Phi}^\mathrm{T}(t\!-\!\tau'') \boldsymbol{\nabla}_{\!\!\mathbf{z}}  \, , \nonumber
\end{align}
which can be simplified to
\begin{align}
\underline{\boldsymbol{\delta\!\mathcal{L}}}(\tau,t) &= \boldsymbol{\Phi}(\tau,t) \, \mathbf{z} + \left[  \boldsymbol{\Phi}(\tau,t) \, \boldsymbol{\sigma}_{\!T}(t,t) - \boldsymbol{\sigma}_{\!T}(\tau,t) \right] \boldsymbol{\nabla}_{\!\!\mathbf{z}} \, ,
\end{align}
where $\boldsymbol{\sigma}_{\!T}(\tau,t)$ is the two-time thermal covariance in Eq.~\eqref{eq:sigmaT}.
As our noise is Gaussian, all operations can be reduced to integrals over the homogeneous propagator and two-time thermal covariance.

%%%%%%%%%%
\subsection{Consistent results} \label{ref:consistent}
\subsubsection{External Forcing of an Oscillator}
Consider the quantum damped oscillator with exact master equation given by Eq.~\eqref{eq:QBML}
under the perturbation of an external force
\begin{eqnarray}
\boldsymbol{\delta\!\mathcal{L}}(t) &=& - \boldsymbol{\nabla}_{\!\!\mathbf{z}}^\mathrm{T} \, \mathbf{F}(t) \, ,
\end{eqnarray}
in the full unitary theory.
It was proven in Ref.~\cite{QBM} that this is not the force one finds in the open-system master equation.
In evaluating the first-order correction, Eq.~\eqref{eq:L1}, we must invoke our rules for transforming system derivatives.
Here we can specifically use Eq.~\eqref{eq:D(tau,t)} to obtain
\begin{align}
\underline{\boldsymbol{\delta\!\mathcal{L}}}(\tau,t) =&\, -\boldsymbol{\nabla}_{\!\!\mathbf{z}}^\mathrm{T} \, \boldsymbol{\Phi}(t\!-\!\tau)  \, \mathbf{F}(\tau) \, , \\
\boldsymbol{\mathcal{L}}_1(t) =&\, \boldsymbol{\delta\!\mathcal{L}}(t) \\
& - \int_0^t \!\! d\tau \left\{ \frac{d}{dt} -\mathrm{Ad}[\boldsymbol{\mathcal{L}}_0(t)] \right\} \boldsymbol{\nabla}_{\!\!\mathbf{z}}^\mathrm{T} \, \boldsymbol{\Phi}(t\!-\!\tau) \, \mathbf{F}(\tau) \, . \nonumber
\end{align}
The diffusion generator in $\boldsymbol{\mathcal{L}}_0(t)$ \eqref{eq:QBML} commutes with the external forcing,
but the homogeneous generator does not commute and one easily finds
\begin{align}
\boldsymbol{\mathcal{L}}_1(t) &= \boldsymbol{\delta\!\mathcal{L}}(t) - \boldsymbol{\nabla}_{\!\!\mathbf{z}}^\mathrm{T} \int_0^t \!\! d\tau \left\{ \frac{d}{dt} + \boldsymbol{\mathcal{H}}(t) \right\} \boldsymbol{\Phi}(t\!-\!\tau) \, \mathbf{F}(\tau) \, .
\end{align}
which is \emph{exactly} the correct effective force;
note that this correction vanishes for local dissipation where $\boldsymbol{\mathcal{H}}(t)$ is constant and $\dot{\boldsymbol{\Phi}}(t) = -\boldsymbol{\mathcal{H}} \, \boldsymbol{\Phi}(t)$.
The true force is not seen in the non-Markovian master equation because, despite its Markovian representation,
the response is still inherently nonlocal.

%%%%%%%
\subsubsection{Linear forcing of a free particle}
Finally we consider a peturbation of linear forces which take the form
\begin{eqnarray}
\boldsymbol{\delta\!\mathcal{L}}(t) &=& \boldsymbol{\nabla}_{\!\!\mathbf{z}}^\mathrm{T} \, \mathbf{K}(t) \, \mathbf{z} \, ,
\end{eqnarray}
so that the perturbative solutions \eqref{eq:phi} to the Langevin equation are simply given by
\begin{eqnarray}
\boldsymbol{\Phi} &=& \boldsymbol{\Phi}_0 + \boldsymbol{\Phi}_1 + \cdots \, , \\
\boldsymbol{\Phi}_1 &=& - \int_0^t \!\! d\tau \, \boldsymbol{\Phi}_0(t\!-\!\tau) \, \mathbf{K}(\tau) \, \boldsymbol{\Phi}_0(\tau) \, . \label{eq:Phi1}
\end{eqnarray}
This is sufficient to calculate the perturbative master equation using the formulas  \ref{sec:QBML}.
The simplest coefficients to calculate are the homogeneous coefficients
\begin{align}
\boldsymbol{\mathcal{H}}(t) &= \boldsymbol{\mathcal{H}}_0(t) + \boldsymbol{\mathcal{H}}_1(t) + \cdots \, , \\
%\boldsymbol{\mathcal{H}}_1 &= \mathbf{K} + \left[ \dot{\boldsymbol{\Phi}}_0 *  \boldsymbol{\mathsf{K}} * \boldsymbol{\Phi}_0 \right] \boldsymbol{\Phi}_0^{-1} + \boldsymbol{\mathcal{H}}_0 \left[ \boldsymbol{\Phi}_0 *  \boldsymbol{\mathsf{K}} * \boldsymbol{\Phi}_0 \right] \boldsymbol{\Phi}_0^{-1} \, , \\
\boldsymbol{\mathcal{H}}_1(t) &= \mathbf{K}(t) \\
& + \int_0^t \!\! d\tau \left\{ \left[ \frac{d}{dt} + \boldsymbol{\mathcal{H}}_0(t) \right] \boldsymbol{\Phi}(t\!-\!\tau) \right\} \mathbf{K}(\tau) \, \boldsymbol{\Phi}(\tau,t) \, , \nonumber
\end{align}
which would be trivial in the limit of local dissipation.
Our new perturbative formalism requires evaluation of the reduced operator
\begin{align}
\underline{\boldsymbol{\delta\!\mathcal{L}}}(\tau,t) &\equiv \left\langle \mathbf{G}_\mathrm{C}(t,\tau) \,  \boldsymbol{\nabla}_{\!\!\mathbf{z}}^\mathrm{T} \, \mathbf{K}(\tau) \, \mathbf{z}  \, \mathbf{G}_\mathrm{C}(\tau,0) \right\rangle_{\mathrm{E}} \left\langle \mathbf{G}_\mathrm{C}(t,0) \right\rangle_\mathrm{E}^{-1} \, ,
\end{align}
which from the results of Sec.~\ref{sec:LinearBack} can be shown to be
\begin{align}
\underline{\boldsymbol{\delta\!\mathcal{L}}}(\tau,t) &=  \boldsymbol{\nabla}_{\!\!\mathbf{z}}^\mathrm{T} \, \boldsymbol{\Phi}(t-\tau) \, \mathbf{K}(\tau) \label{eq:LForce} \\
&\times \left\{ \boldsymbol{\Phi}(\tau,t) \, \mathbf{z} + \left[  \boldsymbol{\Phi}(\tau,t) \, \boldsymbol{\sigma}_{\!T}(t,t) - \boldsymbol{\sigma}_{\!T}(\tau,t) \right] \boldsymbol{\nabla}_{\!\!\mathbf{z}} \right\} \, . \nonumber
\end{align}
It is then a straightforward calculation of Eq.~\eqref{eq:L1} to show that the homogeneous terms here \emph{exactly} reproduce those generated by \eqref{eq:Phi1}.
Comparison of the diffusion coefficients is considerably more taxing, but they also work out to be exactly the same.

%%%%%%%%%%
\subsection{New results} \label{sec:LRules}
From Sec.~\ref{ref:consistent} we now have for the external driving force
\begin{eqnarray}
\boldsymbol{\delta\!\mathcal{L}}(t) &=& - \boldsymbol{\nabla}_{\!\!\mathbf{z}}^\mathrm{T} \, \mathbf{F}(t) \, , \\
\underline{\boldsymbol{\delta\!\mathcal{L}}}(\tau,t) &=& - \boldsymbol{\nabla}_{\!\!\mathbf{z}}^\mathrm{T} \, \boldsymbol{\Phi}(t-\tau) \, \mathbf{F}(t) \, ,
\end{eqnarray}
whereas for the linear force
\begin{eqnarray}
\boldsymbol{\delta\!\mathcal{L}}(t) &=& \boldsymbol{\nabla}_{\!\!\mathbf{z}}^\mathrm{T} \, \mathbf{K}_0(t) \, \mathbf{z} \, , 
%\underline{\boldsymbol{\delta\!\mathcal{L}}}(\tau,t) &=&  \boldsymbol{\nabla}_{\!\!\mathbf{z}}^\mathrm{T} \, \boldsymbol{\Phi}(t-\tau) \, \mathbf{K}_0(\tau) \left\{ 
%\boldsymbol{\Phi}(\tau,t) \, \mathbf{z} + \left[  \boldsymbol{\Phi}(\tau,t) \, \boldsymbol{\sigma}_{\!T}(t,t) - \boldsymbol{\sigma}_{\!T}(\tau,t) \right] \boldsymbol{\nabla}_{\!\!\mathbf{z}}
%\right\} \, .
\end{eqnarray}
we have the two-time operator \eqref{eq:LForce}.
In any case, the first-order master equation perturbative in these forces is given by
\begin{align}
\boldsymbol{\mathcal{L}}_1(t) &= \boldsymbol{\delta\!\mathcal{L}}(t) + \int_0^t \!\! d\tau \left\{ \frac{d}{dt} -\mathrm{Ad}[\boldsymbol{\mathcal{L}}_0(t)] \right\} \underline{\boldsymbol{\delta\!\mathcal{L}}}(\tau,t) \, , \\
&=  \left\{ \frac{d}{dt} -\mathrm{Ad}[\boldsymbol{\mathcal{L}}_0(t)] \right\} \int_0^t \!\! d\tau \, \underline{\boldsymbol{\delta\!\mathcal{L}}}(\tau,t) \, ,
\end{align}
in terms of their respective two-time open-system forces.
These results are compatible with known results for the linear system
and now we will proceed to derive new results.
First we consider an arbitrary quantum deformation
\begin{eqnarray}
\boldsymbol{\delta\!\mathcal{L}}(t) &=& \frac{\partial^d}{\partial p^d} \, F_{db}(t) \, x^b  \, ,
\end{eqnarray}
and then express it in matrix notation as
\begin{eqnarray}
\boldsymbol{\delta\!\mathcal{L}}(t) &=& \left( \boldsymbol{\nabla}_{\!\!\mathbf{z}}^\mathrm{T} \hat{\mathbf{p}} \right)^{\!d} F_{db}(t) \left(\hat{\mathbf{x}}^{\!\mathrm{T}} \mathbf{z} \right)^{\!b}  \, ,
\end{eqnarray}
where $\hat{\mathbf{x}}$ and $\hat{\mathbf{p}}$ are unit vectors in phase space.
Applying the rules of Sec.~\ref{eq:LinearRules}, the two-time operator takes the form
\begin{align}
\underline{\boldsymbol{\delta\!\mathcal{L}}}(\tau,t) =&\, \left[ \boldsymbol{\nabla}_{\!\!\mathbf{z}}^\mathrm{T} \boldsymbol{\Phi}(t-\tau) \, \hat{\mathbf{p}} \right]^{\!d} F_{db}(\tau) \label{eq:dLexp} \\
& \times \sum_{k=0}^b \binom{b}{k} \left[\hat{\mathbf{x}}^{\!\mathrm{T}} \boldsymbol{\Phi}(\tau,t) \, \mathbf{z} \right]^{\!b-k} \boldsymbol{\Delta}^{\![k]}(\tau,t) \, , \nonumber \\
\boldsymbol{\Delta}^{\![k]}(\tau,t) \equiv&\, \left\langle \mathbf{G}_\mathrm{C}(t,0) \left[ \hat{\mathbf{x}}^{\!\mathrm{T}} \!\! \int_0^t \!\! d\tau' \, \boldsymbol{\Phi}_\mathrm{f}(\tau,\tau') \, \boldsymbol{\Xi}(\tau') \right]^{\!k} \right\rangle_{\!\!\!\boldsymbol{\Xi}} \nonumber \\
& \times \left\langle \mathbf{G}_\mathrm{C}(t,0) \right\rangle_{\boldsymbol{\Xi}}^{-1} \, ,
\end{align}
where we have expanded the transformed coordinates via the binomial theorem.
The final step is to evaluate the propagated noise moments via successive application of Novikov's formula.
The zeroth moment is trivial and the first moment we have already calculated
\begin{align}
\boldsymbol{\Delta}^{\![0]}(\tau,t) &= 1 \, , \\
\boldsymbol{\Delta}^{\![1]}(\tau,t) &= \hat{\mathbf{x}}^{\!\mathrm{T}} \left[ \boldsymbol{\Phi}(\tau,t) \, \boldsymbol{\sigma}_{\!T}(t,t) - \boldsymbol{\sigma}_{\!T}(\tau,t) \right] \boldsymbol{\nabla}_{\!\!\mathbf{z}} \, ,
\end{align}
while all higher-order operators can be determined recursively from integration by parts and the product rule.
\begin{align}
\boldsymbol{\Delta}^{\![k]}(\tau,t) =&\, \boldsymbol{\Delta}^{\![1]}(\tau,t) \, \boldsymbol{\Delta}^{\![k-1]}(\tau,t) \\
& - (k\!-\!1) \, \mathbf{s}(\tau,t) \, \boldsymbol{\Delta}^{\![k-2]}(\tau,t) \, , \nonumber \\
\mathbf{s}(\tau,t) \equiv&\, \hat{\mathbf{x}}^{\!\mathrm{T}} \! \left[ \boldsymbol{\sigma}_{\!T}(\tau,\tau) + \boldsymbol{\Phi}(\tau,t) \, \boldsymbol{\sigma}_{\!T}(t,t) \, \boldsymbol{\Phi}^\mathrm{T}(\tau,t) \right]  \hat{\mathbf{x}} \nonumber \\
& - 2 \, \hat{\mathbf{x}}^{\!\mathrm{T}} \! \left[ \boldsymbol{\Phi}(\tau,t) \, \boldsymbol{\sigma}_{\!T}(t,\tau) \right]  \hat{\mathbf{x}} \, ,
\end{align}
and so $\boldsymbol{\Delta}^{\![k]}(\tau,t)$ is a $k^\mathrm{th}$-order differential operator.
Thus we have explicit rules for all master equations with polynomial potentials.

%%%%%%%%%%%%%%
\subsubsection{Quadratic forcing}
A quadratic correction to the spring force (with some cubic correction also in mind) can be written
\begin{eqnarray}
\boldsymbol{\delta\!\mathcal{L}}(t) &=& k_1(t) \left( \frac{\partial}{\partial p} x^2 - \frac{1}{12} \frac{\partial^3}{\partial p^3} \right) \, ,
\end{eqnarray}
where the second term arises only in the quantum Fokker-Planck equation;
it carries an additional dimensional factor of $\hbar^2$ and generates the quantum deformation of trajectories.
To simplify calculation we rewrite these terms in matrix notation as
\begin{align}
\boldsymbol{\delta\!\mathcal{L}}(t) &= k_1(t) \left( \boldsymbol{\nabla}_{\!\!\mathbf{z}}^\mathrm{T} \hat{\mathbf{p}} \right) \left(\hat{\mathbf{x}}^{\!\mathrm{T}} \mathbf{z} \right)^2 - \frac{k_1(t)}{12} \left(\boldsymbol{\nabla}_{\!\!\mathbf{z}}^\mathrm{T} \hat{\mathbf{p}} \right)^3 \, .
\end{align}
Applying expansion formula \eqref{eq:dLexp},
we find the two-time operator to take the form
\begin{align}
\underline{\boldsymbol{\delta\!\mathcal{L}}}(\tau,t) &=
k_1(\tau) \left( \boldsymbol{\nabla}_{\!\!\mathbf{z}}^\mathrm{T} \boldsymbol{\Phi}(t\!-\!\tau) \, \hat{\mathbf{p}} \right) \left(\hat{\mathbf{x}}^{\!\mathrm{T}} \boldsymbol{\Phi}(\tau,t) \, \mathbf{z} \right)^2 \nonumber \\
& + 2 \, k_1(\tau) \left( \boldsymbol{\nabla}_{\!\!\mathbf{z}}^\mathrm{T} \boldsymbol{\Phi}(t\!-\!\tau) \, \hat{\mathbf{p}} \right) \left(\hat{\mathbf{x}}^{\!\mathrm{T}} \boldsymbol{\Phi}(\tau,t) \, \mathbf{z} \right) \boldsymbol{\Delta}^{\![1]}(\tau,t) \nonumber \\
& + k_1(\tau) \left( \boldsymbol{\nabla}_{\!\!\mathbf{z}}^\mathrm{T} \boldsymbol{\Phi}(t\!-\!\tau) \, \hat{\mathbf{p}} \right) \left( \boldsymbol{\Delta}^{\![1]}(\tau,t)^2 - \mathbf{s}(\tau,t) \right) \nonumber \\
& - \frac{k_1(\tau)}{12} \left(\boldsymbol{\nabla}_{\!\!\mathbf{z}}^\mathrm{T} \boldsymbol{\Phi}(t\!-\!\tau) \, \hat{\mathbf{p}} \right)^3 \, . 
\end{align}

%%%%%%%%%
\subsubsection{Cubic forcing}
The cubic correction to the spring force can be written
\begin{eqnarray}
\boldsymbol{\delta\!\mathcal{L}}(t) &=& k_2(t) \left( \frac{\partial}{\partial p} x^3 - \frac{1}{4} \frac{\partial^3}{\partial p^3} x \right) \, ,
\end{eqnarray}
and in matrix notation
\begin{align}
\boldsymbol{\delta\!\mathcal{L}}(t) &= k_2(t) \left( \boldsymbol{\nabla}_{\!\!\mathbf{z}}^\mathrm{T} \hat{\mathbf{p}} \right) \left(\hat{\mathbf{x}}^\mathrm{T} \mathbf{z} \right)^3 - \frac{k_2(t)}{4} \left(\boldsymbol{\nabla}_{\!\!\mathbf{z}}^\mathrm{T} \hat{\mathbf{p}} \right)^3 \left(\hat{\mathbf{x}}^\mathrm{T} \mathbf{z} \right) \, ,
\end{align}
which yields the two-time operator
\begin{align}
& \underline{\boldsymbol{\delta\!\mathcal{L}}}(\tau,t) =
k_2(\tau) \left( \boldsymbol{\nabla}_{\!\!\mathbf{z}}^\mathrm{T} \boldsymbol{\Phi}(t\!-\!\tau) \, \hat{\mathbf{p}} \right) \left(\hat{\mathbf{x}}^{\!\mathrm{T}} \boldsymbol{\Phi}(\tau,t) \, \mathbf{z} \right)^3 \nonumber \\
& + 2 \, k_2(\tau) \left( \boldsymbol{\nabla}_{\!\!\mathbf{z}}^\mathrm{T} \boldsymbol{\Phi}(t\!-\!\tau) \, \hat{\mathbf{p}} \right) \left(\hat{\mathbf{x}}^{\!\mathrm{T}} \boldsymbol{\Phi}(\tau,t) \, \mathbf{z} \right)^2 \boldsymbol{\Delta}^{\![1]}(\tau,t) \nonumber \\
& + 2 \, k_2(\tau) \left( \boldsymbol{\nabla}_{\!\!\mathbf{z}}^\mathrm{T} \boldsymbol{\Phi}(t\!-\!\tau) \, \hat{\mathbf{p}} \right) \left(\hat{\mathbf{x}}^{\!\mathrm{T}} \boldsymbol{\Phi}(\tau,t) \, \mathbf{z} \right)  \left( \boldsymbol{\Delta}^{\![1]}(\tau,t)^2 - \mathbf{s}(\tau,t) \right)  \nonumber \\
& + k_2(\tau) \left( \boldsymbol{\nabla}_{\!\!\mathbf{z}}^\mathrm{T} \boldsymbol{\Phi}(t\!-\!\tau) \, \hat{\mathbf{p}} \right) \left( \boldsymbol{\Delta}^{\![1]}(\tau,t)^3 - 3\, \mathbf{s}(\tau,t) \, \boldsymbol{\Delta}^{\![1]}(\tau,t) \right) \nonumber \\
& - \frac{k_2(\tau)}{4} \left(\boldsymbol{\nabla}_{\!\!\mathbf{z}}^\mathrm{T} \boldsymbol{\Phi}(t\!-\!\tau) \, \hat{\mathbf{p}} \right)^3 \left(\hat{\mathbf{x}}^\mathrm{T} \boldsymbol{\Phi}(\tau,t) \, \mathbf{z} \right) \nonumber \\
& - \frac{k_2(\tau)}{4} \left(\boldsymbol{\nabla}_{\!\!\mathbf{z}}^\mathrm{T} \boldsymbol{\Phi}(t\!-\!\tau) \, \hat{\mathbf{p}} \right)^3  \boldsymbol{\Delta}^{\![1]}(\tau,t) \, . 
\end{align}

%%%%%%%%%%%
%\section{Local Damping}
%Here we calculate the necessary thermal correlations for linear perturbations and approximately local damping,
%$\tilde{\gamma}(\varepsilon) \approx \tilde{\gamma}_0$ but with some very high cutoff frequency for which the kernel vanishes.
%These calculations can also be generalized to the case of rational damping following the results of Ref.~\cite{QBM}.
%For local damping the homogeneous coefficients $\boldsymbol{\mathcal{H}}(t)$ are constant in time and the homogeneous propagator is simply
%\begin{eqnarray}
%\boldsymbol{\Phi}(t,\tau) &=& e^{-(t-\tau)\boldsymbol{\mathcal{H}}} \, .
%\end{eqnarray}
%Combining equations \eqref{eq:sigmaT}, \eqref{eq:nuq}, and \eqref{eq:nu},
%the two-time thermal correlation can be expressed
%\begin{equation}
%\boldsymbol{\sigma}_T(t_1,t_2) = \frac{m}{2\pi}\! \int_{-\infty}^{+\infty} \!\!\! d\varepsilon \, \tilde{\gamma}(\varepsilon) \, \tilde{\kappa}(\varepsilon) \int_0^{t_1}\!\!\!d\tau_1  \int_0^{t_2}\!\!\!d\tau_2 \, \cos[\varepsilon(\tau_1\!-\!\tau_2)] \, e^{-(t_1-\tau_1) \boldsymbol{\mathcal{H}}} \mtiny{ \left[ \begin{array}{cc} 0 & 0 \\ 0 & 1 \end{array} \right] } e^{-(t_2-\tau_2) \boldsymbol{\mathcal{H}}^\mathrm{T}} \, ,
%\end{equation}
%in terms of the FDR kernel
%\begin{eqnarray}
%\tilde{\kappa}(\varepsilon) &\equiv&  \varepsilon \coth\!\left( \frac{\varepsilon}{2T} \right) \, .
%\end{eqnarray}

\section{Discussion}
We have derived a fairly general theory of strong-coupling perturbation for continuous variable systems.
Our formalism makes heavy use of the phase-space representation as
the zeroth-order problem is a linear one and the phase-space representation is perhaps the simplest formalism for that case.
Essentially we are perturbing off of QBM, where the QBM-like model provides us with stochastic coordinates which do not solve the problem,
but integrate the environment dynamics.
This results in a combined nonlinear and stochastic dynamics of the unraveled system.
From there we apply the standard perturbation theory of master equations, one wherein the perturbative time-translation generator is calculated.
This method retains important Lie-group symmetries and grants the possibility of non-secular behavior (thus allowing for late-time solutions).
As one might expect our master equation is highly non-Markovian and highly model specific.

These result are given in a form which is calculable, but far more complicated than what one would desire at first order.
At least two avenues of attack must be considered.
In the larger view one must question if there is a more suitable formalism which includes all necessary ingredients: the stochastic map, open and closed-system propagation, etc..
The linearity of the zeroth-order dynamics is simplest here in the phase-space representation, but perhaps we could trade some of that simplicity for
a representation which is more apt for the nonlinearity.
Or perhaps a combined approach could best utilize properties of different representations.
On the other hand, in the more localized view one must address some simplified calculation of the noise averages
which predominantly rely upon knowledge of the two-time thermal correlation and integrals thereof.
We have demonstrated that a true strong-coupling master equation exists, and we have explicitly given its form,
but a large amount of work remains to apply our results to representative physical setups.

\bibliography{bib}{}
\bibliographystyle{apsrev}

\end{document}